\documentclass[useAMS,usenatbib,usegraphicx]{mn2e}
\usepackage{soul}
\usepackage{ulem}
\def\la{\raise.5ex\hbox{$<$}\kern-.8em\lower 1mm\hbox{$\sim$}}
\def\ga{\raise.5ex\hbox{$>$}\kern-.8em\lower 1mm\hbox{$\sim$}}
\def\be{\begin{equation}}
\def\ee{\end{equation}}
\def\ba{\begin{eqnarray}}
\def\ea{\end{eqnarray}}

\def\Omegastar{\Omega_\ast}
\def\OmegaK{\Omega_{\mathrm{K}}}
\def\Mdotin{\dot{M}_{\mathrm{in}}}
\def\Mdot{\dot{M}}

\def\Pdot{\dot{P}}
\def\Pddot{\ddot{P}}
\def\Msun{M_{\odot}}

\def\rin{r_{\mathrm{in}}}
\def\rlc{r_{\mathrm{LC}}}
\def\rout{r_{\mathrm{out}}}
\def\rco{r_{\mathrm{co}}}

\def\Lx{L_{\mathrm{x}}}
\def\Md{M_{\mathrm{d}}}

\def\rA{r_{\mathrm{A}}}

\def\Tp{T_{\mathrm{p}}}

\def\dM*{\delta M_*}

\def\P0min{P_{0,{\mathrm{min}}}}

\def\p{\propto}

\def\fast{\omega_{\ast}}
\def\Alfven{Alfv$\acute{e}$n}
\def\418{SGR 0418+5729}
\def\142{AXP 0142+61}

\def\Pzero{P$_{\mathrm{0}}$ }
\def\Bzero{$B_{\mathrm{0}}$}

\title[On The Evolution of The Radio Pulsar PSR J1734$-$3333]{On The Evolution of The Radio Pulsar PSR J1734$-$3333}
\author[\c{S}. \c{C}al\i\c{s}kan et al.]{\c{S}. \c{C}al\i\c{s}kan$^{1}$\thanks{E-mail:
scaliskan@sabanciuniv.edu}, \"{U}. Ertan$^{1}$, M. A. Alpar$^{1}$, J. E. Tr\"{u}mper$^{2}$, and N. D. Kylafis$^{3}$\\
$^{1}$Sabanc\i\ University, Orhanl\i - Tuzla, \.Istanbul, 34956, Turkey\\
$^{2}$Max-Planck-Institut f\"{u}r extraterrestrische, Physik Geissenbachstra\ss{}e, 85748 Garching bei M\"{u}nchen, Germany\\
$^{3}$University of Crete, Physics Department \& Institute of Theoretical \& Computational Physics, 71003 Heraklion, Crete, Greece}
\begin{document}

\pagerange{\pageref{firstpage}--\pageref{lastpage}} \pubyear{2002}
\maketitle

\label{firstpage}

\begin{abstract}
Recent measurements showed that the period derivative of the `high-B' radio pulsar PSR J1734$-$3333 is increasing with time. For neutron stars evolving with fallback disks, this rotational behavior is expected in certain phases of the long-term evolution. Using the same model as employed earlier to explain the evolution of anomalous X-ray pulsars and soft gamma-ray repeaters, we show that the period, the first and second period derivatives and the X-ray luminosity of this source can simultaneously acquire the observed values for a neutron star evolving with a fallback disk. We find that the required strength of the dipole field that can produce the source properties is in  the range of $ 10^{12} - 10^{13}$ G on the pole of the neutron star. When the model source reaches the current state properties of PSR J1734$-$3333, accretion onto the star has not started yet, allowing the source to operate as a regular radio pulsar. Our results imply that PSR J1734$-$3333 is at an age of $\sim 3 \times 10^4 - 2 \times 10^5$ years. Such sources will have properties like the X-ray dim isolated neutron stars or transient AXPs at a later epoch of weak accretion from the diminished fallback disk.
\end{abstract}


\section{Introduction}

The discovery of several new classes of isolated neutron stars, namely the
 anomalous X-ray pulsars (AXPs) and soft gamma-ray-burst repeaters (SGRs), 
 the X-ray dim isolated neutron stars (XDINs), the compact central objects in certain
  supernova remnants (CCOs), and the rotating radio transients (RRATs), has
   brought into focus the question of possible evolutionary links \citep[see, e.g.,][]{Kaspi10, Popov08}. The existence of radio pulsars with large 
    inferred dipole magnetic moments, close to, and in fact partly overlapping
     with, the range of magnetar fields inferred for AXPs and SGRs ($\ga$ 
     10$^{14}$ G), and the observation of radio pulses from some AXPs and SGRs, 
     further highlight the possibility of links and raise questions about the 
     similarities and differences among these sources. The locations and 
     evolutionary tracks of pulsars in the $P-\dot{P}$ diagram hold the keys to 
     deciphering the links. The recent measurement of the peculiar braking 
     index $n = 0.9 \pm 0.2$  of PSR J1734$-$3333, the lowest among the measured braking indices of young pulsars, with period $P$ = 1.17 s and period derivative 
     $\dot{P}$ = 2.28 $\times 10^{-12}$ s s$^{-1}$ \citep{Espinoza11} is an 
     exciting new clue. 

Analysing the $P-\dot{P}$ diagram of all isolated pulsars in terms of evolution by rotationally powered dipole radiation into the vacuum, starting with the initial rotation rate and the magnetic dipole moment at birth, and assuming that the dipole moment remains constant, fails to explain the distribution of all young pulsars on the $P-\dot{P}$ diagram or to shed  light on the possible connections of the distinct classes. Evolution of the magnetic-dipole moment and its angle with the rotation axis is one direction of extending the picture, as suggested in the work reporting the braking index of PSR J1734$-$3333 \citep{Espinoza11}. Magnetic-field evolution and amplification to lead to magnetar strength $10^{14}-10^{15}$ G surface fields have been  invoked to explain the bursts of SGRs and AXPs \citep{DT92}. This does not require that the magnetar strength fields are necessarily in the $dipole$ component that controls the spin-down of the pulsar and thereby its evolution on the $P-\dot{P}$ diagram. Indeed, the recently published $\dot{P}$ upper limit of SGR 0418+5729 \citep{Rea10} has shown that the surface dipole field of this pulsar is at most 7.5 $\times 10^{12}$ G at the equator (1.5 $\times 10^{13}$ G at the poles). This has been interpreted in terms of the decay of the magnetar dipole moment of this SGR \citep{Turolla11}.
All measured braking indices \citep{Becker09} deviate from the n = 3 value, characteristic for dipole radiation in vacuum. Braking indices n $<$ 3 imply a growing magnetic dipole moment (or growth of the dipole moment component perpendicular to the rotation axis). Thus, to understand the behavior of pulsars and magnetars in different parts of the $P-\dot{P}$ diagram requires growth, or decay, of the magnetic dipole moment perpendicular to the rotation axis, at rates depending on the sources.

An alternative avenue for a more general picture of isolated pulsar evolution is to allow for the possibility of interaction with matter around the star, so that the emission is not dipole radiation in vacuum. An effective possibility is that some matter left over from the supernova explosion is actually bound to the neutron star, in the form of a `fallback' accretion disk, as it necessarily carries angular momentum, as proposed by \citet*{CHN00} to explain the properties of AXPs. \citet{Alpar01}  suggested that the presence or absence, and the initial mass, of a fallback disk could be the third initial condition, complementing the initial rotation rate and dipole magnetic moment, to determine the subsequent evolution of different classes of neutron stars. A first simple application of this idea to pulsar braking indices and motion across the $P-\dot{P}$ diagram was presented by \citet*{AAY01}. Discoveries of radio pulsars with long periods and large period derivatives suggested that these sources could have evolutionary links with AXPs and SGRs (\citealt{Kaspi10, Espinoza11}; see, e.g., \citealt{Mereghetti08} for a recent review of AXPs and SGRs).

In the present paper we apply the fallback disk model in detail to the evolution of PSR J1734$-$3333 and show that the model can explain all the properties of this source, including its braking index. 
In Section 2 we outline the model and examine evolutionary tracks for a neutron star with a fallback disk to search for  scenarios leading to the present properties ($P, \Pdot, \Pddot$, and $\Lx$) of PSR J1734--3333.  We  trace all possible initial conditions, namely the initial period, dipole field, and disk mass, that can produce the source properties. We discuss the results of our model calculations in Section 3, and summarise our conclusions in Section 4.

\section{The Model}

In the fallback disk model (FDM) \citep{CHN00,Alpar01} the period and luminosity evolution are determined by the interaction between the neutron star and a fallback disk around it. The mid-infrared detections from the AXPs 4U 0142+61 \citep*{Wang06} and 1E 2259+586 \citep{Kaplan09} are consistent with the presence of fallback disks around these sources \citep{EEEA07}. It was shown by \citet{EC06} that the observed near-IR luminosities and the upper limits of AXPs/SGRs are compatible with the expectations of FDM. The values of the dipole-field strength at the pole of the neutron star \Bzero, indicated by FDM fits to optical/IR data of 4U 0142+61 \citep{EEEA07} and by the results of the work explaining the long-term $P, \Pdot$, and X-ray luminosity, $\Lx$,  evolution of AXPs and SGRs, are less than $10^{13}$ G in all cases \citep{EEEA09, AEC11}.  It was proposed in these papers that the magnetar strength fields needed to power the bursts must be residing in quadrupole and higher multipole components. The higher multipole fields fall off with distance from the star more rapidly than the dipole field does, leaving the dipole field to determine the interaction with the disk and the resulting torques.

The interactive evolution of the neutron star and the fallback disk can have epochs with accretion as well as radio pulsar epochs. The neutron star enters the accretion regime and experiences an efficient torque if and when the inner edge of the fallback disk penetrates into the light cylinder.  The neutron star can then spin down to long periods of several seconds on timescales from $\sim 10^{3}$  to $\sim 10^5$ yr,  depending on the disk torques, the dipole-field strength $B_0$ and the disk mass $\Md$. 
The first (ejector) phase of evolution without accretion could last from several years to more than $10^5$ yr depending on  $B_0$, $\Md$, and initial period $P_0$. During this phase, the neutron star is a radio pulsar. 
In the present work, we show that PSR J1734$-$3333 is likely to be in the radio pulsar phase without accretion at present and that the accretion epoch could start at a future time.  In the accretion phase, at a time depending on the initial parameters, the inner disk will reach the light cylinder and the accretion will stop. After the accretion phase, the disk could remain attached to the light cylinder and the disk torque could still remain active while its efficiency gradually decreases to the level of the dipole radiation torque. 
Unlike a steady-state disk in a binary, where the accreting stage can be sustained on the evolutionary timescales of the binary and the companion, the fallback disk around an isolated neutron star will diminish. 
From the beginning, the outermost parts of the disk are always at low temperatures. Eventually, temperature even in the inner disk becomes too low to sustain viscosity. The disk then becomes passive, mass inflow and disk torques terminate.
The decreasing luminosity and disk torque together lead to the observed period clustering \citep[see][for details]{EEEA09, AEC11}.

For PSR J1734$-$3333, we investigate the evolution mainly in the initial radio pulsar phase. 
We address the peculiar braking index of PSR J1734$-$3333, thereby exploring the effect of a possible fallback disk on the evolution of isolated radio pulsars across the $P - \Pdot$  diagram. The recent measurement of the second derivative of the period, $\Pddot$ = $5.3 \times 10^{-24}$ s s$^{-2}$  \citep{Espinoza11}, provides an opportunity to test FDM  evolutionary scenarios more stringently than before, checking for the first time for simultaneous agreement of $\Pddot$ with $P, \Pdot$, and X-ray luminosity, $\Lx$. The X-ray luminosity of the source is $7.3 \times 10^{31} - 6.6 \times 10^{32}$ erg s$^{-1}$, taking into account the 25\% error margins for the distance estimate \citep{Olausen12}. It is interesting that the $\Pddot$ of PSR J1734$-$3333 is positive. This means that the pulsar is evolving towards the upper right corner, the AXP/SGR region, of the $P - \Pdot$ diagram. For an isolated neutron star evolving by magnetic-dipole radiation in vacuum, this would require a dipole field growing in time. Note that the toroidal and  dipole fields of AXPs/SGRs, starting from the early phase of evolution, should $decrease$ rather rapidly with time in the magnetar model \citep[see, e.g.,][]{Turolla11}.

The model we employ for PSR J1734$-$3333 is the same as the one we used to investigate the long term evolution of AXPs/SGRs in our earlier work. The details of the model are described in \citet{EEEA09} and \citet{AEC11}. We start with an initially extended disk with an inner radius equal to the \Alfven ~radius  
\be
\rA = (G M)^{-1/7}~\mu^{4/7}~ \Mdotin^{-2/7},
\ee
where $\Mdotin$ is the rate of mass flow arriving at the inner disk, $G$ is the gravitational constant, $M$ and $\mu$ are the mass and the magnetic-dipole moment of the neutron star. When the inner disk radius $\rin$, calculated by Equation 1, exceeds the light cylinder radius $\rlc$, we set $\rin = \rlc$ \citep{AEC11}. This assumes that the inner disk remains linked on the closed field lines when it cannot enter the light cylinder. In this phase, accretion is not possible, and pulsed radio emission is allowed. In the phase of spin-down with accretion, when the inner disk is inside the light cylinder and greater than the co-rotation radius ($\rco < \rin < \rlc$) , a fraction of the matter could be propelled from the system while the remaining fraction accretes onto the neutron star.

In the accretion phase, we calculate the disk torque acting on the neutron star using 
\be 
N =  \frac{1}{2} ~\Mdotin ~(G M \rin)^{1/2}~ (1-\fast^2) = I~ \dot{\Omegastar}
\ee  
\citep{EE08}, where $I$ is the moment of inertia of the neutron star. The fastness parameter is defined as $\fast = \Omegastar/\OmegaK(\rin)$, where  $\OmegaK(\rin)$ is the angular frequency of the disk at  $\rin = \rA$ and $\Omegastar$ is the angular frequency of the neutron star. 
Using Equations 1 and 2, it is found that $\Pdot \p B^2 $, independent of $\Mdot$  and  $P$ when the system is not close to rotational equilibrium. This indicates that $\Pdot$ is constant and $\Pddot = 0$ in this phase. When a high-luminosity AXP/SGR is approaching (or receding from) rotational equilibrium $\Pddot$ is negative (positive). 
In the early radio phase, the magnetic dipole radiation torque could dominate the disk torque for fast born pulsars ($P_0~\la$ 50 ms) for $B_0 \sim 10^{12}$ G. We calculate the total torque including also the magnetic dipole torque 
$N_{dip} = -2\mu^{2}\Omegastar^{3} / 3c^{3}$.

In our model calculations, we find that over the spin history of PSR J1734$-$3333, the dipole radiation torque remains 2-3 orders of magnitude weaker than the disk torque for an initial period $P_0 \sim$ 300 ms.  For a given field strength, the ratio of the torques depends on the chosen disk torque model and the initial period P$_{0}$. For instance, our results agree with the results of \citet*{Yan12}, who found that for slow-born pulsars (with P$_{0}$ $\sim$ 300 ms) the disk torque dominates for the first $\sim$ 10$^{5}$ years. In their model, for fast-born pulsars (P$_{0}$ $\sim$ 5 ms), the dipole torque is the dominant mechanism for the first $\sim$ 10$^{5}$ years. The disk torque we employ in our model is more efficient than that used in \cite{Yan12}, and depending on the disk mass, it could start to dominate the dipole torque in an earlier phase of the evolution for fast-born pulsars (which probably represent a small set among newborn neutron stars, lying in the tail of a Gaussian distribution with mean 300 ms and standard deviation 150 ms, according to the simulations of \citealt{FGK06}).

The mass flow rate at the inner disk is calculated at each evolutionary step by solving the diffusion equation with an initially thin disk surface density profile \citep[e.g.,][]{EEEA09}. Initially, for numerical reasons, we set the outer disk radius at $\rout = 5 \times 10^{14}$ cm. After the first time step, $\rout$  has a dynamical evolution such that the temperature at $\rout$  remains equal to $\Tp$, and beyond this radius the disk is assumed to be in a viscously inactive phase. Because of decreasing irradiation flux, temperatures and $\rout$ also decrease gradually with time. The initial disk mass is found by integrating the initial surface density profile. Since the position of the initial outer disk radius is not well known, our $\Md$ values may not reflect the actual full disk mass.
In this phase, we substitute $\rin = \rlc$ to calculate the disk torque (Equation 2). When the dipole torque is negligible, we find $\Pdot \p \Mdotin  P^{7/2}$ in this phase. We perform numerical calculations to follow the evolution of $\Mdotin$ together with corresponding $P, \Pdot$, and $\Pddot$ at each time step \citep{EEEA09, AEC11}.  We repeat the calculations until we identify the initial conditions that can produce the observed $P, \Pdot, \Pddot$ and also, in the present case, $\Lx$ of PSR J1734$-$3333 simultaneously, at an age when the disk does not penetrate the light cylinder, allowing for radio pulsar activity.  

In the radio pulsar phase, for a neutron star with the properties of PSR J1734$-$3333 the source of the X-ray luminosity is very likely to be the cooling luminosity. For the fast born pulsars, the dipole radiation luminosity could remain well above the cooling luminosity until the periods increase to a few 100 ms. We include the cooling and dipole luminosities in addition to dissipation due to magnetic and disk torques \citep{Alpar07} in the calculation of the total luminosity in the radio phase. 
In the long-term evolution of AXPs/SGRs and XDINs, when the sources reach long periods of a few seconds, the presence of cooling luminosity extends the life time of the disk by providing irradiation even in the absence of accretion luminosity. During the initial radio phase, it does not have a significant effect on the evolution, but, through the effect of irradiation on the disk evolution and disk torques, it affects the time at which the model source acquires the observed rotational properties ($P, \Pdot, \Pddot$) simultaneously. In our model, we employ the intrinsic cooling luminosity evolution calculated by \citet*{Page06} with the assumption of an isothermal neutron star with radius 12 km and mass 1.4 $\Msun$. The results are not sensitive to the choice among standard cooling scenarios.

\begin{figure}
\includegraphics[width=0.8\textwidth,angle=270]{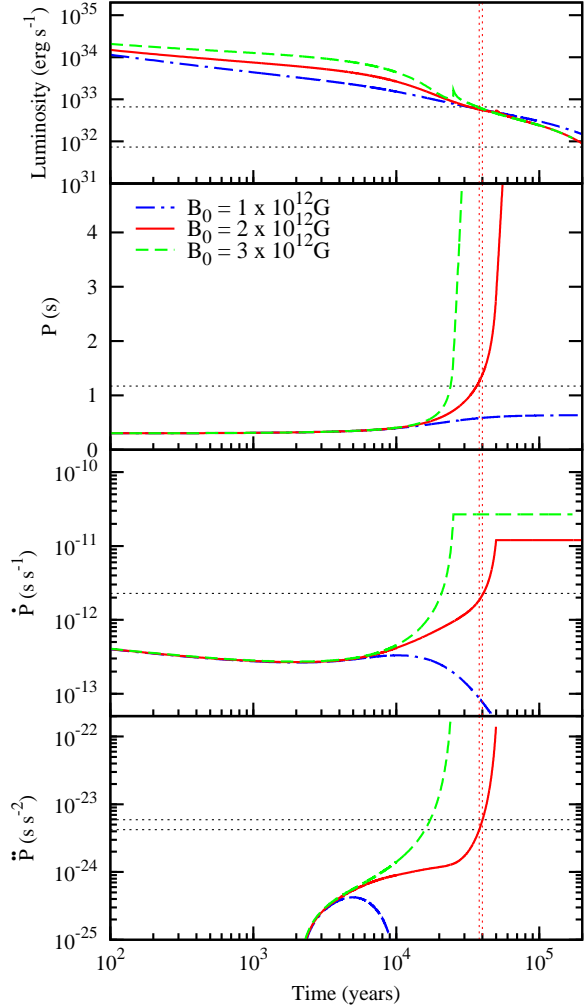}
\caption{Evolution of the luminosity, period, first and second period derivatives of the model sources. Horizontal lines show the properties of PSR J1734$-$3333, with the observational uncertainties in $\Lx$ and $\Pddot$. The vertical lines are to show the time period over which the solid (red) model curve traces the uncertainty range of $\Pddot$.  Values of  $B_0$ are given in the second panel. For these calculations, we have taken $\Md = 3 \times 10^{-7} \Msun$ and $P_0 = 300$ ms. In the accretion phase, sources enter the constant  $\Pdot$ phase and $\Pddot$ becomes 0 (see text for details). }
\end{figure}

\begin{figure}
\includegraphics[width=0.8\textwidth,angle=270]{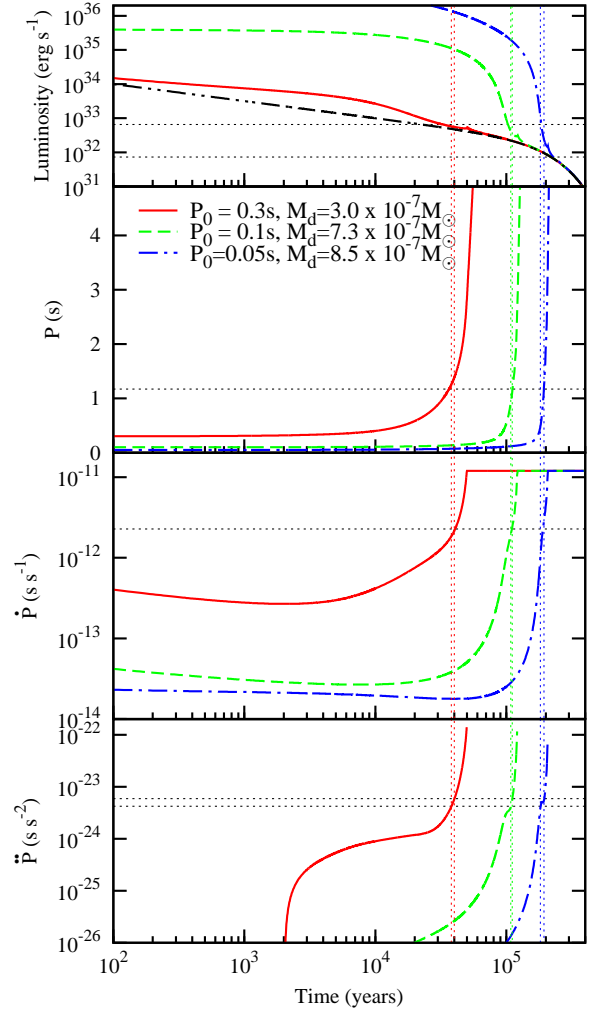}
\caption{Evolution of the luminosity, period, first and second period derivatives of model sources. Horizontal dotted lines represent the properties of PSR J1734$-$3333  with the range of  uncertainties in $\Lx$ and $\Pddot$. These illustrative model curves are obtained for $B_0 = 2 \times 10^{12}$ G. The values of initial period and disk mass are given in the second panel. The cooling luminosity is shown with the dot-dot-dashed (black) curve. It is seen that the source properties could be well reproduced with different initial periods. Between the vertical lines given with the same color, the model sources trace the uncertainty range of $\Pddot$  (see text for details).  }
\end{figure}

\begin{figure}
\includegraphics[width=0.8\textwidth,angle=270]{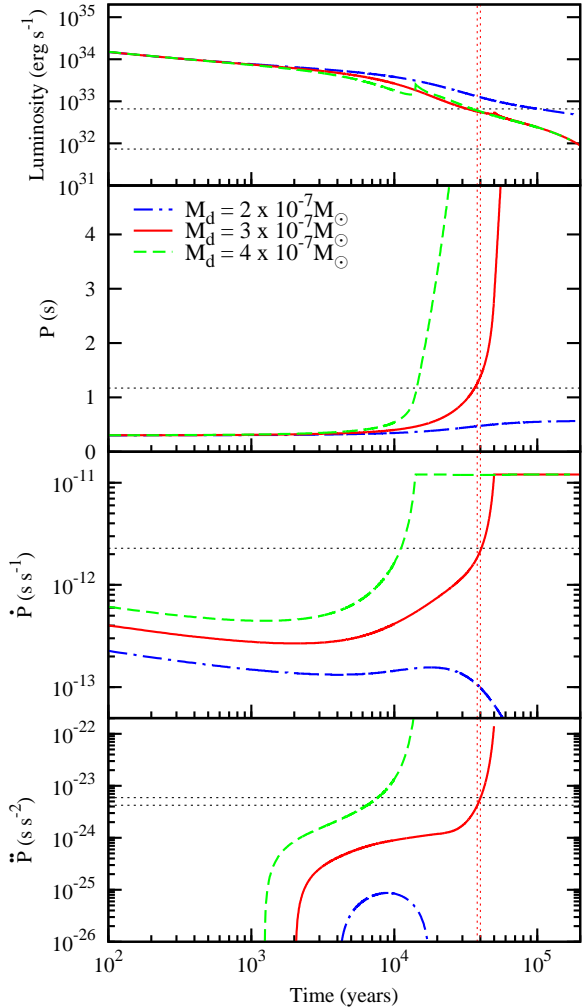}
\caption{Evolution of the luminosity, period, first and second period derivatives, for models that do not work for PSR J1734--3333. Horizontal dotted lines show the properties of PSR J1734$-$3333, with the uncertainties in $\Lx$ and $\Pddot$.  These model curves are obtained with $B_0 = 2 \times 10^{12}$ G  and $P_0 = 300$ ms. The solid lines are the same as the solid lines in Figures 1 and 2, for the model that works. We also present two illustrative  model curves for  smaller and greater $\Md$ that cannot represent the evolution of PSR J1734$-$3333  (see text for details).  }
\end{figure}

\begin{figure}
\includegraphics[width=0.8\textwidth,angle=270]{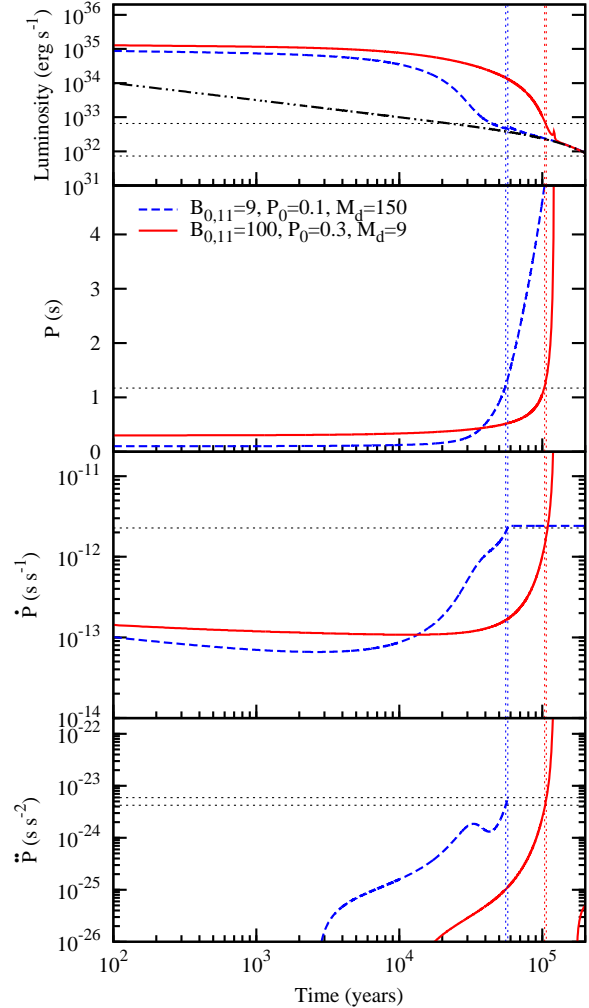}
\caption{
Evolution of the luminosity, period, first and second period derivatives, for model sources with the lowest and highest allowed $B_0$ values. The magnetic field is given in units of 10$^{11}$ G, the initial period is in seconds and the disk mass is given in units of 10$^{-8} \Msun$. The cooling luminosity is shown with the dot-dot-dashed (black) curve. }
\end{figure}

\begin{figure}
\includegraphics[width=0.44\textwidth,angle=270]{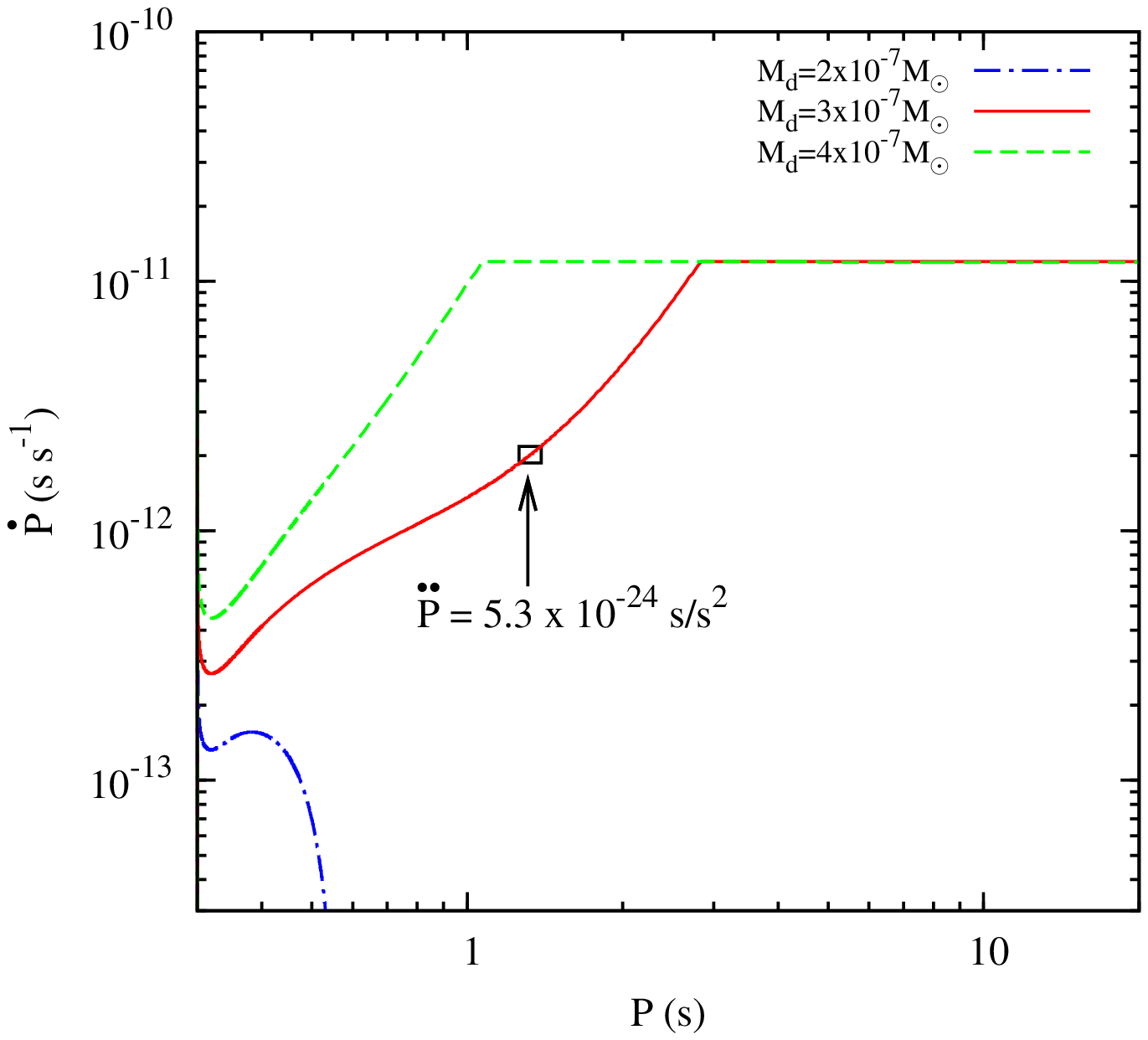}
\caption{The evolution of the three model sources of Figure 3 on the $P - \Pdot$~diagram. The values of initial disk mass $\Md$ are shown in the Figure. All sources start with \Pzero = 300 ms. The model source with the lowest $\Md$ (blue) never enters the accretion regime and its period converges to $\sim$ 0.5 s. The model source with the highest $\Md$ (green) enters the accretion regime early on and after 10$^{5}$ years it has P $>$ 30 s. The model represented by the solid curve can reproduce the properties of PSR J1734$-$3333  (see Figure 3). The rectangle shows the current position of the source on the $P - \Pdot$~diagram. The size of the rectangle represents the uncertainty in the measured $\Pddot$~value. }
\end{figure}

\section{Results and Discussion}

The model curves that can account for the properties of PSR J1734$-$3333 are seen in Figures 1 - 4. Performing many simulations tracing the initial conditions,  we obtain reasonable results for the range of disk masses $\Md = 9.0 \times 10^{-8} - 1.5 \times 10^{-6} {\Msun}$ and for dipole fields with $B_0 = 9 \times 10^{11} - 1 \times 10^{13} $ G.  The model is in very good agreement with $P, \Pdot, \Pddot$, and $\Lx$ of the source with $B_0 \simeq 2 \times 10^{12}$ G (solid curve in Figure 1). 
For the solid curve in Figure 1, the disk mass $\Md = 3 \times 10^{-7} \Msun$ and the initial period $P_0 = 300$ ms.
In Figure 1, we also present illustrative model curves (dashed and dot-dashed) that cannot represent the evolution of PSR J1734$-$3333. 
For the viable model in Figure 1, and all three viable models shown in Figure 2, the source is powered by intrinsic cooling when the observed properties of PSR J1734$-$3333 are attained.
For this radio pulsar, considering all reasonable model curves, the simultaneous agreement with all these model parameters leads to a model estimate of the present age $\sim 3 \times 10^{4} - 2 \times 10^{5}$ years.
At an epoch later than the present age of PSR J1734$-$3333, the disk eventually penetrates the light cylinder. Accretion then starts, and the star enters a constant $\Pdot$ phase. This is also observed as a small abrupt rise in the luminosity curve. There is no substantial  increase  in the luminosity when accretion starts, because the mass-flow rate has already decreased to low levels by the time the inner disk penetrates the light cylinder. Reasonable model curves indicate that accretion will start at a future time of order $\sim 10^4$ yr from the present. For other model source histories, accretion can start at earlier times and the accretion luminosity can be orders of magnitude greater than the cooling luminosity. Evolutionary curves for such models do not simultaneously produce all the observed properties of PSR J1734$-$3333.

Figure 1 shows that $\Pddot$ reaches values of order $\sim 2 \times 10^{-22}$ s s$^{-2}$ at the end of the radio pulsar epoch. Before the accretion phase, from t $\sim 10^3$ years to the present age, the braking index for model sources varies from $\sim 5$ to $\sim -1$.  In the accretion phase, $\Pddot = 0$, $\Pdot$ becomes constant and the braking index remains $\sim$ 2 when the source is not very close to rotational equilibrium.     

The mass accretion from the disk starts at a time that depends on the initial period $P_0$, as well as on $B_0$ and $\Md$ \citep{EEEA09}.  A large range of $P_0$ values are allowed for producing the source properties. In Figure 2, for a given dipole field ($B_0 = 2 \times 10^{12}$ G), we present three illustrative model curves  with initial periods of 50, 100 and 300 ms.  To obtain acceptable models, smaller $P_0$ values require greater $\Md$. All model curves given in Figure 2 produce the X-ray luminosity and the rotational  properties of PSR J1734$-$3333 at different epochs. In the early radio-pulsar phase, the source of the luminosity is either the intrinsic cooling or magnetic dipole radiation of the neutron star, depending on $B_0$ and $P_0$. Reasonable model curves indicate that current luminosity of PSR J1734$-$3333 is produced by the intrinsic cooling luminosity while the source is slowing down by the disk torques. We find that the rates of dissipation inside the star due to dipole and disk torques \citep{Alpar07} do not contribute significantly to the total luminosity in the radio-pulsar phase of this source at present. 
A part of the dipole luminosity is emitted in the X-ray band. For model sources that have low $P_0$ and/or high $B_0$ values, the dipole radiation could be the dominant luminosity in the early phases of evolution. For instance, for initially fast-rotating model sources with $B_0= 2 \times 10^{12}$ G, the luminosity of magnetic dipole radiation dominates the intrinsic cooling luminosity until the period increases to beyond $\sim$ 200 ms (see Figure 2). For $B_0~\ga~10^{13}$ G, the dipole luminosity remains above the upper limit of the observed luminosity when the rotational properties of PSR J1734$-$3333 are acquired. We note that the blackbody nature of the X-ray spectrum \citep{Olausen12} also indicates that the source of the observed luminosity is likely to be the intrinsic cooling rather than magnetic dipole radiation.
For the model sources with $B_0~\la~9 \times 10^{11}$ G,  accretion starts before the current rotational properties of PSR J1734$-$3333 are reached. 
To sum up, the luminosity, $P$, $\Pdot$ and $\Pddot$ values of PSR J1734$-$3333 can be acquired by the model sources only for $B_0$ values in $\sim 9 \times 10^{11} - 1 \times 10^{13}$ G range.
The two model curves with the minimum and maximum allowed values of $B_0$ and $\Md$ are given in Figure 4.

The important parameters that affect the long-term evolution of the model sources are the initial period $P_0$, the minimum temperature of the active disk $\Tp$, which is degenerate with the irradiation efficiency $C$, \citep[but $C$ is restricted by optical and IR observations of AXPs;][]{EC06}, the disk mass $\Md$, and the dipole magnetic field strength $B_0$ on the surface of the star. In our earlier work and in the present paper, we take $\Tp \simeq$ 100 $-$ 200 K. Since $\Tp$ together with $C$ determine the end of the active disk phase, they have almost no effect on the evolution during the initial radio phase. The initial period can affect the time of onset of accretion, but similar long term evolutionary curves could be found for a large range of initial periods. To follow the evolution of model sources in the initial radio phase, there remain only two important parameters: $\Md$ and $B_0$. We repeat the simulations tracing $\Md$ and $B_0$ to find the entire allowed ranges of these parameters that can produce the properties of PSR J1734$-$3333 simultaneously.  

Our results  indicate that PSR J1734$-$3333 could be at an age in the range  $\sim 3 \times 10^{4} - 2 \times 10^{5}$ yr depending on the actual initial conditions of the star. For $B_0 = 2 \times 10^{12}$ G and $P_0 = 300$ ms, $\Md$ = $3 \times 10^{-7} \Msun$ gives the observed properties of PSR J1734$-$3333 simultaneously, while for $P_0 = 100$ ms the disk mass is $\Md$ = $7.3 \times 10^{-7} \Msun$ and for $P_0 = 50$ ms $\Md$ = $8.5 \times 10^{-7} \Msun$. 
For $B_0 = 2 \times 10^{12}$ G and $P_0 = 300$ ms, illustrative model curves obtained with different disk masses are given in Figure 3.  The solid (red) curve is the same as the tracks given in Figures 1 and 2 with the same $B_0$ and $P_0$. It is seen that the model with the greatest $\Md$ starts the accretion phase earlier, while the source with the smallest  $\Md$ never enters the accretion phase. These models, represented by dashed and dot-dashed curves in Figure 3, cannot produce all present properties of PSR J1734$-$3333 simultaneously. 

In later phases of  evolution, radio pulsars similar to PSR J1734$-$3333 are likely to be observed either as transient AXPs in the accretion phase, if they are detected by means of soft gamma-ray bursts, or as  XDINs after the termination of the accretion phase, while the disk torques are still effective. We also present the evolution of these model sources on the $P - \Pdot$ diagram in Figure 5. A paper on the evolution of XDINs is in preparation \citep{EAC13}.

Our long-term evolution model of neutron stars with fallback disks differs from earlier fallback disk models described in \cite{CHN00} and \citet*{Menou01a}, as well as \cite{Alpar01} and \cite{AAY01}  in two crucial ingredients of the models: (1) the minimum temperature ($\Tp$) at which the viscous activity stops and the disk becomes passive, and (2) the torque expression employed in the models, which determine the  $\Mdot$ dependence and spin evolution of the neutron star.

According to the fallback disk model proposed by \cite{Menou01a}, the disk becomes inactive at accretion rates at which the disk cools through the thermal viscous instability. To explain the braking indices of young neutron stars with high X-ray luminosities, \citet*{Menou01b} employed the same model. In this model, the minimum mass-flow rate for the disk to remain active and interact with the magnetic field is a few times 10$^{14}$ g s$^{-1}$. The properties of the transient AXPs, discovered later with inferred quiescent accretion rates much lower than the minimum critical limit of \cite{Menou01a}, seem hard to be explained together with persistent sources in a single coherent picture.   Another difficulty arises when applying this model to PSR J1734$-$3333 and to other pulsars with low luminosities. At present, the inner disk of PSR J1734$-$3333 cannot enter the light cylinder, and therefore the pulsed radio emission is allowed, provided that the mass-flow rate of the disk is less than about 10$^{13}$ g s$^{-1}$ (even for a relatively low dipole field strength of 10$^{12}$ G). With this mass-flow rate, the disk is not active in this model. To put it in other words, the model used in \cite{Menou01a} cannot be applied to PSR J1734$-$3333. A similar problem emerges in the fallback disk model used in \cite{CHN00}, since the disk is assumed to make a transition to ADAF regime at mass inflow rates even higher than the minimum critical rates given in \cite{Menou01a}.  In \cite{CHN00} the disk is assumed to have no contribution to the torque in the radio phase, that is, this model cannot be applied to PSR J1734$-$3333 either. In our model we do not encounter this problem: the disk becomes passive at low temperatures $\Tp \sim$ 100 $-$ 200 K, in accordance with theoretical work by \cite{Inutsuka05}, starting from the outermost disk as described in detail in \cite{EEEA09} and \cite{AEC11}.

The evolutionary model curves of the rotational properties of  neutron stars and their dependence on mass flow rate of the disk are sensitive to the disk torque models employed in the calculations. The motivation leading us to the particular torque model employed in earlier work \citep{EE08, EEEA09, AEC11} and in the present work was the peculiarity of the contemporaneous X-ray luminosity and period evolution of the transient AXP XTE J1810-197 in the X-ray enhancement/outburst phase.
Nevertheless, the observations of  XTE J1810-197 \citep{Gotthelf07} show that in the range of accretion rates from a few 10$^{13}$ to 10$^{15}$ g s$^{-1}$, the torque remains almost independent of mass-inflow rate. We use the same torque model that is in good agreement with this observation, in all subsequent work on spin evolution of neutron stars with fallback disks.  This behaviour, observed in XTE J1810-197, cannot be accounted for by earlier disk torque models.

\section{Conclusions}

We have shown through detailed analysis that the observed period, the first and second  period derivatives, and the X-ray luminosity of PSR J1734$-$3333 can be simultaneously reached by a neutron star evolving with a fallback disk. The model is compatible with the pulsed radio emission of the source, since the present source properties are reached at a time when the  accretion of matter from the disk is not allowed yet. 

The observed properties of PSR J1734$-$3333 can be obtained at an age of $\sim 3 \times 10^4 - 2 \times 10^5$ yr, with a range of initial periods and with disk masses ($9 \times 10^{-8}\Msun  < \Md <  1.5 \times 10^{-6}\Msun $), relatively low compared to those estimated for AXPs/SGRs.  Dipole fields with strength in the range $9 \times 10^{11} - 1 \times 10^{13}$ G, together with  appropriate disk masses and initial periods, give reasonable model curves. 

PSR J1734$-$3333 has not been observed in IR to the best of our knowledge. A search for IR emission from the fallback disk in PSR J1734$-$3333 will be of great interest. The infrared emission from the disk depends on several factors, like inclination angle of the disk, irradiation efficiency, outer and inner disk radii. We will present elsewhere the expected IR emission from PSR J1734$-$3333 and sources similar to it in detail, together with a discussion of the assumed parameters.

We expect that PSR J1734$-$3333 will evolve into the accretion phase within another $\sim 10^4$ yr (see Figures 1 and 2). By the onset of the accretion phase, the mass-flow rate of the inner disk will have decreased to a very low level such that accretion does not significantly contribute to the total X-ray luminosity. Radio pulsars following evolutionary curves similar to that of PSR J1734$-$3333 could be detected in the accretion phase if they show soft gamma-ray bursts like AXPs/SGRs - these would be identified as transient AXPs. Such sources are not likely to emit radio waves, having spun down to long periods by this late stage of evolution. Only if they are sufficiently close to Earth we could detect and identify them as XDINs. Some of these sources could evolve to periods longer than the AXP/SGR periods \citep{EEEA09}.  It is hard to detect them due to low X-ray luminosities. In short, sources like PSR J1734$-$3333 may evolve to become transient AXPs or XDINS. The evolution of XDINs and their connection with AXPs/SGRs will be the subject of our future work.   

\section*{Acknowledgments}
We thank the anonymous referee for his/her useful comments. We acknowledge research support from Sabanc\i\ University, from
T\"{U}B{\.I}TAK (The Scientific and Technical Research Council of
Turkey) through grant 110T243. M.A.A. is a member of the Science Academy, Istanbul, Turkey. We thank K. Yavuz Ek\c{s}i and M. Hakan Erkut for useful discussions.

\label{lastpage}


\begin{thebibliography}{99}
\bibitem[\protect\citeauthoryear{Alpar}{2001}]{Alpar01} Alpar M. A., 2001, ApJ, 554, 1245.
\bibitem[\protect\citeauthoryear{Alpar}{2007}]{Alpar07} Alpar M. A., 2007, Ap\&SS, 308, 133.
\bibitem[\protect\citeauthoryear{Alpar, Ankay \& Yazgan}{Alpar et al.}{2001}]{AAY01} Alpar M. A., Ankay A., Yazgan E., 2001, ApJ, 557, L61
\bibitem[\protect\citeauthoryear{Alpar, Ertan \& \c{C}al\i\c{s}kan}{Alpar et al.}{2011}]{AEC11} Alpar M.~A., Ertan \"{U}., \c{C}al\i\c{s}kan \c{S}., 2011, ApJ, 732, L4
\bibitem[\protect\citeauthoryear{Becker}{2009}]{Becker09} Becker W., 2009, in "Neutron Stars and Pulsars", W. Becker, ed., Astrophysics and Space Science Library 357, 91
\bibitem[\protect\citeauthoryear{Chatterjee, Hernquist \& Narayan}{Chatterjee et al.}{2000}]{CHN00} Chatterjee P., Hernquist L., Narayan R., 2000, ApJ, 534, 373
\bibitem[\protect\citeauthoryear{Duncan \& Thompson}{1992}]{DT92} Duncan R. C., Thompson C., 1992, ApJ, 392, L9
\bibitem[\protect\citeauthoryear{Ertan \& \c{C}al{\i}\c{s}kan}{2006}]{EC06} Ertan \"{U}., \c{C}al{\i}\c{s}kan \c{S}., 2006, ApJ, 649, L87
\bibitem[\protect\citeauthoryear{Ertan \& Erkut}{2008}]{EE08} Ertan \"{U}., Erkut M. H., 2008, ApJ, 673, 1062
\bibitem[\protect\citeauthoryear{Ertan et al.}{2007}]{EEEA07} Ertan \"{U}., Erkut M. H., Ek\c si K. Y., Alpar M. A., 2007, ApJ, 657, 441
\bibitem[\protect\citeauthoryear{Ertan et al.}{2009}]{EEEA09} Ertan \"{U}., Ek\c si K. Y., Erkut M. H., Alpar M. A., 2009, ApJ, 702, 1309
\bibitem[\protect\citeauthoryear{Ertan et al.}{2013}]{EAC13} Ertan \"{U}. et al., 2013, in preparation
\bibitem[\protect\citeauthoryear{Espinoza et al.}{2011}]{Espinoza11} Espinoza C.~M., Lyne A.~G., Kramer M., Manchester R.~N., Kaspi V.~M., 2011, ApJ, 741, L13
\bibitem[\protect\citeauthoryear{Faucher-Gigu{\`e}re \& Kaspi}{2006}]{FGK06} Faucher-Gigu{\`e}re, C.-A., \& Kaspi, V.~M.\ 2006, ApJ, 643, 332
\bibitem[\protect\citeauthoryear{Gotthelf \& Halpern}{2007}]{Gotthelf07} Gotthelf, E.~V., Halpern, J.~P., 2007, Ap\&SS, 308, 79
\bibitem[\protect\citeauthoryear{Inutsuka \& Sano}{2005}]{Inutsuka05} Inutsuka, S., Sano, T., 2005, ApJ, 628, L155
\bibitem[\protect\citeauthoryear{Kaplan et al.}{2009}]{Kaplan09} Kaplan D.~L., Chakrabarty D., Wang Z., Wachter S., 2009, ApJ, 700, 149
\bibitem[\protect\citeauthoryear{Kaspi}{2010}]{Kaspi10} Kaspi V.~M., 2010, PNAS, 107, 7147 
\bibitem[\protect\citeauthoryear{Menou, Perna \& Hernquist}{Menou et al.}{2001a}]{Menou01a} Menou, K., Perna, R., Hernquist, L. 2001a,  ApJ, 559, 1032
\bibitem[\protect\citeauthoryear{Menou, Perna \& Hernquist}{Menou et al.}{2001b}]{Menou01b} Menou, K., Perna, R., Hernquist, L. 2001b,  ApJ, 554, L063
\bibitem[\protect\citeauthoryear{Mereghetti}{2008}]{Mereghetti08} Mereghetti S., 2008,  A\&AR, 15, 225
\bibitem[\protect\citeauthoryear{Olausen et al.}{2012}]{Olausen12} Olausen, S.~A., Zhu, W.~W., Vogel, J.~K., et al., 2012, arXiv:1211.5387 
\bibitem[\protect\citeauthoryear{Page, Geppert \& Weber}{Page et al.}{2006}]{Page06}Page, D., Geppert, U., \& Weber, F.\ 2006, Nuclear Physics A, 777, 497 
\bibitem[\protect\citeauthoryear{Popov}{2008}]{Popov08} Popov S. B., 2008, Physics of Particles and Nuclei, 39, 1136
\bibitem[\protect\citeauthoryear{Rea et al.}{2010}]{Rea10} Rea N. et al., 2010, Science, 330, 944
\bibitem[\protect\citeauthoryear{Turolla et al.}{2011}]{Turolla11} Turolla R., Zane S., Pons J.~A., Rea N., 2011, ApJ, 740, 105
\bibitem[\protect\citeauthoryear{Wang, Chakrabarty \& Kaplan}{Wang et al.}{2006}]{Wang06} Wang Z., Chakrabarty D., Kaplan D.~L., 2006, Nature, 440, 772
\bibitem[\protect\citeauthoryear{Yan, Perna \& Soria}{Yan et al.}{2012}]{Yan12} Yan, T., Perna, R., Soria, R., 2012, MNRAS, 423, 2451
\end{thebibliography}
\end{document}